\documentclass[aps,prd,preprint,nofootinbib]{revtex4}
\usepackage{graphicx}

\begin{document}
\title{GENXICC2.0: An Upgraded Version of the Generator for Hadronic
Production of Double Heavy Baryons $\Xi_{cc}$, $\Xi_{bc}$ and
$\Xi_{bb}$}
\author{Chao-Hsi Chang$^{1,2,3}$\footnote{zhangzx@itp.ac.cn}
Jian-Xiong Wang$^{4}$\footnote{jxwang@ihep.ac.cn} and Xing-Gang
Wu$^{1}$\footnote{wuxg@cqu.edu.cn} }
\address{$^1$Department of Physics, Chongqing University, Chongqing 400044,
China \\
$^2$CCAST (World Laboratory), P.O.Box 8730, Beijing 100080, China\\
$^3$Institute of Theoretical Physics, Chinese Academy of Sciences,
Beijing 100190, China\\
$^4$Institute of High Energy Physics, P.O.Box 918(4), Beijing
100049, P.R. China}
\date{\today}

\begin{abstract}
An upgraded (second) version of the package GENXICC ({\bf A
Generator for Hadronic Production of the Double Heavy Baryons
$\Xi_{cc}$, $\Xi_{bc}$ and $\Xi_{bb}$ by C.H. Chang, J.X. Wang and
X.G. Wu,} [its first version: in Comput. Phys. Commun. {\bf 177}
(2007) 467-478]) is presented. Users, with this version being
implemented in PYTHIA and a GNU C compiler, may simulate full events
of the production in various experimental environments conveniently.
In comparison with the previous version, in order to implement it in
PYTHIA properly, a subprogram for the fragmentation of the produced
double heavy diquark to the relevant baryon is complemented and the
interphase of the generator to PYTHIA is changed accordingly. In the
subprogram, with explanation, certain necessary assumptions
(approximations) are made so as to conserve the momenta and the QCD
`color' flow for the fragmentation. \\

\end{abstract}

\maketitle

\noindent{\bf NEW VERSION PROGRAM SUMMARY} \\

\noindent{\it Title of program} : GENXICC2.0 \\

\noindent{\it Program obtained from} : CPC Program Library or the
Institute of Theoretical Physics, Chinese Academy of Sciences,
Beijing, P.R. China: {$www.itp.ac.cn/\,\widetilde{}\;zhangzx/genxicc2.0$}.\\

\noindent{\it Reference to original program} : GENXICC\\

\noindent{\it Reference in CPC} : Comput. Phys. Commun. {\bf 177},
467(2007)\\

\noindent{\it Does the new version supersede the old program}:
No\\

\noindent{\it Computer} : Any LINUX based on PC
with FORTRAN 77 or FORTRAN 90 and GNU C compiler as well.\\

\noindent{\it Operating systems} : LINUX.\\

\noindent{\it Programming language used} : FORTRAN 77/90.\\

\noindent{\it Memory required to execute with typical data} : About
2.0 MB.\\

\noindent{\it No. of bytes in distributed program, (including
PYTHIA6.4)} : About 1.5 MB.\\

\noindent{\it Distribution format} : .tar.gz .\\

\noindent{\it Nature of physical problem} : Hadronic production of
double heavy baryons $\Xi_{cc}$, $\Xi_{bc}$ and $\Xi_{bb}$.\\

\noindent{\it Method of solution} : The code is based on NRQCD
framework. With proper option, it can generate weighted and
un-weighted events of hadronic double heavy baryon production. When
the hadronizations of the produced jets and double heavy diquark are
taken into account in the production, the upgraded version with
proper interface to PYTHIA can well generate the events in full.\\

\noindent{\it Restrictions on the complexity of the problem} : The
color flow, particularly, in the piece of programming the
fragmentation from the produced colorful double heavy diquark into a
relevant double heavy baryon, is treated carefully so as to
implement it in PYTHIA properly. \\

\noindent{\it Reasons for new version} : Responding to the feedback
from users\cite{D0}, we improve the generator mainly by careful
completing the `final nonperturbative process' i.e. the formulation
of the double heavy baryon from relevant intermediate diquark. In
the present version, the information for fragmentation about
momentum-flow and the-color flow, that is necessary for PYTHIA to
generate full events, is retained although reasonable approximations
are made. In comparison with the original version, the upgraded one
can implement it in PYTHIA properly to do the full event simulation
of the double heavy baryon production. \\

\noindent{\it Typical running time} : It depends on which option is
chosen to match PYTHIA when generating the events in full and also
on which mechanism is chosen to generate the events. Typically, for
the most complicated case with gluon-gluon fusion mechanism to
generate the mixed events via the intermediate diquark in
$(cc)[^3S_1]_{\bar{3}}$ and $(cc)[^1S_0]_6$ states, and to generate
1000 events, it takes about 20 hours on a 1.8 GHz Intel P4-processor
machine if IDWTUP=1, whereas to generate $10^6$ events it takes
about 40 minutes only if IDWTUP=3. \\


\noindent{\it Summary of the changes (improvements)} : 1) We try to
explain the treatment of the momentum distribution of the process
more clearly than the original version, and show how the final
baryon is generated through the typical intermediate diquark
precisely.  2) We present color flow of the involved processes
precisely and the corresponding changes for the program are made.
3). The corresponding changes of the program are explained in the
paper.

\section{momentum distribution of the production}

\begin{table}
\begin{center}
\caption{All considered mechanisms for step A, which are defined by
the two parameters {\bf mgenxi} and {\bf ixiccstate}. Here the
symbol gg-mechanism stands for the gluon-gluon fusion mechanism and
etc..} \vskip 0.6cm
\begin{tabular}{|c||c|c|c|}
\hline ~~---~~ & ~~mgenxi=1~~& ~~mgenxi=2~~ & ~~mgenxi=3~~ \\
\hline\hline ~~ixiccstate=1~~ & gg-mechanism, $(cc)_{\bf
\bar{3}}(^3S_1)$ & gg-mechanism, $(bc)_{\bf \bar{3}}(^3S_1)$
& gg-mechanism, $(bb)_{\bf \bar{3}}(^3S_1)$ \\
\hline ~~ixiccstate=2~~ & gg-mechanism, $(cc)_{\bf 6}(^1S_0)$
& gg-mechanism, $(bc)_{\bf 6}(^1S_0)$ & gg-mechanism, $(bb)_{\bf 6}(^1S_0)$ \\
\hline ~~ixiccstate=3~~ & gc-mechanism, $(cc)_{\bf \bar{3}}(^3S_1)$
& gg-mechanism, $(bc)_{\bf 6}(^3S_1)$ & ~~---~~ \\
\hline ~~ixiccstate=4~~ & gc-mechanism, $(cc)_{\bf 6}(^1S_0)$
& gg-mechanism, $(bc)_{\bf \bar{3}}(^1S_0)$ & ~~---~~ \\
\hline ~~ixiccstate=5~~ & cc-mechanism, $(cc)_{\bf \bar{3}}(^3S_1)$
& ~~---~~ & ~~---~~ \\
\hline ~~ixiccstate=6~~ & cc-mechanism, $(cc)_{\bf 6}(^1S_0)$
& ~~---~~ & ~~---~~ \\
\hline
\end{tabular} \label{xi}
\end{center}
\end{table}

In fact, in the program we divide the production of a double heavy
baryon $\Xi_{cc}$ or $\Xi_{bc}$ or $\Xi_{bb}$ into two steps: Step-A
is up-to the production of a relevant double heavy diquark and
Step-B is followed for the fragmentation of the double heavy diquark
into the desired baryon. Therefore, in Step-A, there are three
possible mechanisms: the gluon-gluon fusion mechanism ($g+g$),
gluon-charm collision mechanism ($g+c$) and charm-charm collision
mechanism ($c+c$), so in the program we need to fix one of them to
produce the diquarks $(cc)$, $(bc)$ and $(bb)$ accordingly in term.
All the three mechanisms and the calculation techniques for them are
described in Refs.\cite{genxicc,cqww,cmqw}. In Step-B, it is for the
fragmentation of the double heavy diquark into the desired baryon.
In the step we assume the intermediate diquark is to `decay' into
the baryon plus soft parton(s) exclusively, e.g., either
$(QQ')[^3S_1]_{\bar{3}}\to \Xi_{QQ'}+\bar{q}$ or
$(QQ')[^1S_0]_{\bar{6}}\to \Xi_{QQ'}+\bar{q}+g$ (here $Q,Q'$ denote
$c,b$-quark, $\bar{q}$ an light anti-quark, $g$ a gluon). For
convenience, in the program we name the mechanisms and intermediate
diquarks in terms of the parameters as those in TAB.\ref{xi}, where
{\bf mgenxi} stands for the double-heavy diquarks, $(cc)$ or $(bc)$
or $(bb)$, and {\bf ixiccstate} stands for the mechanisms. Note that
in the previous version of GENXICC (we call it GENXICC1.0), we did
not program how the diquark forms the relevant baryon, but
alternatively we simply assumed that the relevant baryon is formed
with 100\% efficiency.

According to QCD confinement, the produced diquarks $(QQ')$, i.e.
$(cc)$, $(bc)$, $(bb)$, must be fragmented into relevant baryons by
grabbing a light quark $q$ (even suitable number of gluons $g$) with
definite probability. Since the fragmentation for the heavy diquarks
is absent from the available version of PYTHIA, thus in the upgraded
version of GENXICC we program the fragmentation precisely and make
its interphase still to suit PYTHIA properly. For consistency, in
the upgraded version for the fragmentation we adopt the assumptions
and method similar to those taken by PYTHIA \cite{pythia} in the
case for the fragmentation of a color-octet component $(c\bar{c})_8$
into a colorless charmonium. Namely here the double heavy quark to
grab a light quark (with gluons if necessary) from the `environment'
to form a colorless double heavy baryon with a relative possibility
for various flavors of the light quark as $u :d :s :c \simeq
1:1:0.3:10^{-11}$. Hence in the program we introduce three new
parameters {\bf ratiou} (default=1), {\bf ratiod}(default=1), {\bf
ratios}(default=0.3) so as to dictate the probability for a double
heavy diquark to grab a light quark (antiquark) in forming the
relevant baryon finally. These parameters may be changed by setting
the values of the parameters in the parameter.F, when the relative
possibilities for various flavors are assumed precisely. One more
parameter {\bf nbound} is naturally introduced to dictate which type
of baryon: $\Xi_{cc}^{+,++}$ or $\Omega_{cc}^{+}$ ({\bf nbound}=1),
$\Xi_{bc}^{+,0}$ or $\Omega_{bc}^{0}$ ({\bf nbound}=2),
$\Xi_{bb}^{0,-}$ or $\Omega_{bb}^{-}$ ({\bf nbound}=3) is to be
generated from the relevant produced diquark. ({\bf nbound}=4) is to
derive the diquark results that can be generated by previous version
(GENXICC1.0). The relative possibilities for the baryons
$\Xi_{cc}^{+}$, $\Xi_{cc}^{++}$ and $\Omega_{cc}^{+}$ are decided by
the value of {\bf ratiou}, {\bf ratiod} and {\bf ratios}. More
precisely, if the diquark $(cc)[^3S_1]_{\bar{3}}$ is produced, then
it will fragment into $\Xi_{cc}^{++}$ with $43\%$ probability,
$\Xi_{cc}^{+}$ with $43\%$ probability and $\Omega_{cc}^{+}$ with
$14\%$ probability accordingly, when default values of {\bf ratiou},
{\bf ratiod} and {\bf ratios} are taken.

Below we shall only take the hadronic production of $\Xi_{cc}$ as an
example to show how the generator GENXICC works, because the
production for the baryon $\Xi_{bc}$ or $\Xi_{bb}$ is similar.

At the end of the Step-A, the final particles' momenta are set by
using the {\bf phase$\_$gen} routine, that is based on RAMBO (Random
Momentum Booster) program \cite{rambo}, and the irrelevant phase
space is integrated by VEGAS \cite{vegas}. For the Step-B, we adopt
the `decay' method (as that to deal with the intermediate color
octet $(c\bar{c})$ states to produce $J/\psi$ in PYTHIA
\cite{pythia}). According to the method, we start with assuming the
diquark mass to be slightly bigger than that of the baryon
($>2m_c$), so the diquark may `decay' into a relevant baryon by
emitting very soft partons (anti-quark and/or gluons), and the soft
partons take away very tiny momentum from the diquark. To keep gauge
invariance of the hard process in Step-A, we must set $(cc)$-diquark
mass to be $2m_c$ exactly \cite{cqww,cmqw}, thus in the present
case, we set the diquark mass to be $2m'_c$ i.e. slightly bigger
value, e.g. $m'_c\geq m_c+m_q/2$ ($m_q$ is mass of the light quark),
and let the mass of the produced baryon be less than but very close
to $2m_c+m_q$ approximately. In the program, we introduce the
parameters {\bf slqmass} to stand for the soft-light-quark mass
$m_q$(q=u,d,s): {\bf qmassu} for $u$-quark, {\bf qmassd} for
$d$-quark and {\bf qmasss} for $s$-quark respectively. Since the
Step-B in the case for the diquark $(cc)[^3S_1]_{\bar{3}}$ being
produced is of a ($1\to 2$)-body process ($(cc)[^3S_1]_{\bar{3}}\to
\Xi_{cc}+\bar{q}$) and we have averaged the polarization of the
produced diquark $(cc)[^3S_1]_{\bar{3}}$ in Step-A, so we reasonably
treat the fragmentation as that, for the double heavy baryon and the
light anti-quark, the absolute values of the momenta is fixed by
momentum conservation and the masses of the three relevant
particles, whereas the direction of the momenta is isotropic in the
diquark rest frame. It can be found that such a light anti-quark
indeed affects very slightly in the momentum distribution of the
generated double heavy baryon. We should note here that even though
such a soft light quark gives slightly correction to the
distribution in momentum, it must be involved so as to make the
color flow of the process right for each event in full. If the color
flow of the process were not correct, then the program would not be
able to implement in PYTHIA to generate full events. As for the
Step-B in the case for the diquark $(cc)[^1S_0]_{6}$ being produced,
that is of a ($1\to 3$)-body process ($(cc)[^1S_0]_{6}\to
\Xi_{cc}+\bar{q}+g$, one more gluon is needed) indeed due to the
color-flow request as pointed out in the next section, we still
treat it as a ($1\to 2$)-body process for momentum distribution of
the fragmentation by letting $\bar{q}$ and $g$ always move together.
``letting $\bar{q}$ and $g$ always move together'' is a very strong
constraint but here $\bar{q}$ and $g$ are very `light' and carry the
momenta very tiny, so it does not affect the distribution of the
baryon, what we are interested in, substantially.

To record all the useful data, we improve the data manipulation
method. All the recorded data are still put in the directory {\it
data}, while in this directory, we introduce 12 subdirectories; {\it
xiccu}, {\it xiccd}, {\it xiccs}, {\it xibcu}, {\it xibcd}, {\it
xibcs}, {\it xibbu}, {\it xibbd}, {\it xibbs}, {\it ccdiq}, {\it
bcdiq} and {\it bbdiq}. These subdirectories record the needed data
respective to the settings determined by the user in {\bf
parameter.F}, e.g. {\it xiccu}, {\it xiccd} and {\it xiccs} record
data for $\Xi^{++}_{ccu}$, $\Xi^{+}_{ccd}$, $\Omega^{+}_{ccs}$
respectively; {\it ccdiq}, {\it bcdiq} and {\it bbdiq} record data
for $(cc)$-diquark, $(bc)$-diquark, $(bb)$-diquark respectively. We
should note that the subdirectories {\it ccdiq}, {\it bcdiq} and
{\it bbdiq} record the data just as those generated by GENXICC1.0.

\section{the color flow of the production}

Ignoring the color flow of the fragmentation at all, still one can
obtain incomplete information of the event for the baryon
production, such as the observables: total cross section, transverse
momentum $p_t$ and rapidity $y$ differential distributions of the
baryon etc. Thus, even GENXICC1.0, the previous version, can do all
these jobs well. The present (upgraded) version deals with the color
flow not only for the double heavy diquark production but also for
the fragmentation carefully, so by complementing the upgraded
version of the generator into PYTHIA properly, it can generate full
events and record all information of the outgoing partons, besides
that of the baryon itself.

The color flow for $\Xi_{cc}$ and $\Xi_{bb}$ production is much more
involved than that of $\Xi_{bc}$ production, so at the present,
firstly we focus the color flow for $\Xi_{cc}$ production precisely.
Moreover to meet the needs in most cases, we only include the most
important gluon-gluon fusion mechanism for the $\Xi_{bc}$ and
$\Xi_{bb}$ production. The interesting readers can obtain the color
flow for the $\Xi_{bc}$ and $\Xi_{bb}$ production by following the
same method, but we note that for $\Xi_{bc}$ production since the
heavy quarks inside it are different flavored so its production via
the diquark $(bc)$ with the quantum number as
$(bc)[^1S_0]_{\bar{3}}$ or $(bc)[^3S_1]_{\bar{3}}$ or
$(bc)[^1S_0]_6$ or $(bc)[^3S_1]_6$ (four possibilities) instead of
$(QQ)[^3S_1]_{\bar{3}}$ or $(QQ)[^1S_0]_6$ ($Q=c,b$, two
possibilities) for $\Xi_{QQ}$ production.

\subsection{The color flow for $(cc)[^3S_1]_{\bar{3}}$ being produced}

To show the color flow of $\Xi_{cc}$ production, the mechanisms for
$(cc)[^3S_1]_{\bar{3}}$ production and the followed fragmentation
can be more precisely written as
\begin{eqnarray} && gg \to
(cc)[^3S_1]_{\bar{3}}+\bar{c}\bar{c} \to \Big(\Xi^+_{cc}
+\bar{d}\Big) +\bar{c}\bar{c} \;\;{\it or}\;\; \Big(\Xi^{++}_{cc}
+\bar{u}\Big) +\bar{c}\bar{c} \;\;{\it or}\;\;
\Big(\Omega^{+}_{cc} +\bar{s}\Big) +\bar{c}\bar{c}\\
&& gc\to (cc)[^3S_1]_{\bar{3}}+\bar{c} \to
\Big(\Xi^{+}_{cc}+\bar{d}\Big) +\bar{c} \;\;{\it or}\;\;
\Big(\Xi^{++}_{cc} +\bar{u}) +\bar{c}\;\;{\it or}\;\;
\Big(\Omega^{+}_{cc} +\bar{s}) +\bar{c} \\
&& cc\to (cc)[^3S_1]_{\bar{3}}+g \to \Big(\Xi^{+}_{cc}+\bar{d}\Big)
+g \;\;{\it or}\;\; \Big(\Xi^{++}_{cc} +\bar{u}\Big) +g\;\;{\it
or}\;\; \Big(\Omega^{+}_{cc} +\bar{s}\Big) +g
\end{eqnarray}
According to the discussion in Ref.\cite{majp}, for the
$(cc)$-diquark in $^3S_1$ state with the color ${\bf \bar 3}$, one
of the two heavy $c$ quarks emits a soft gluon and then this gluon
splits into a light $q \bar{q}$ pair, and the $(cc)$-pair can
combine the light quark $q$ to form the baryon $\Xi_{cc}$ (in
$|ccq\rangle$ Fock state). To fit the needs of PYTHIA running, here
we program the second step properly, i.e. to deal with the color
flow of the fragmentation process carefully.

\begin{figure}
\centering
\includegraphics[width=0.3\textwidth]{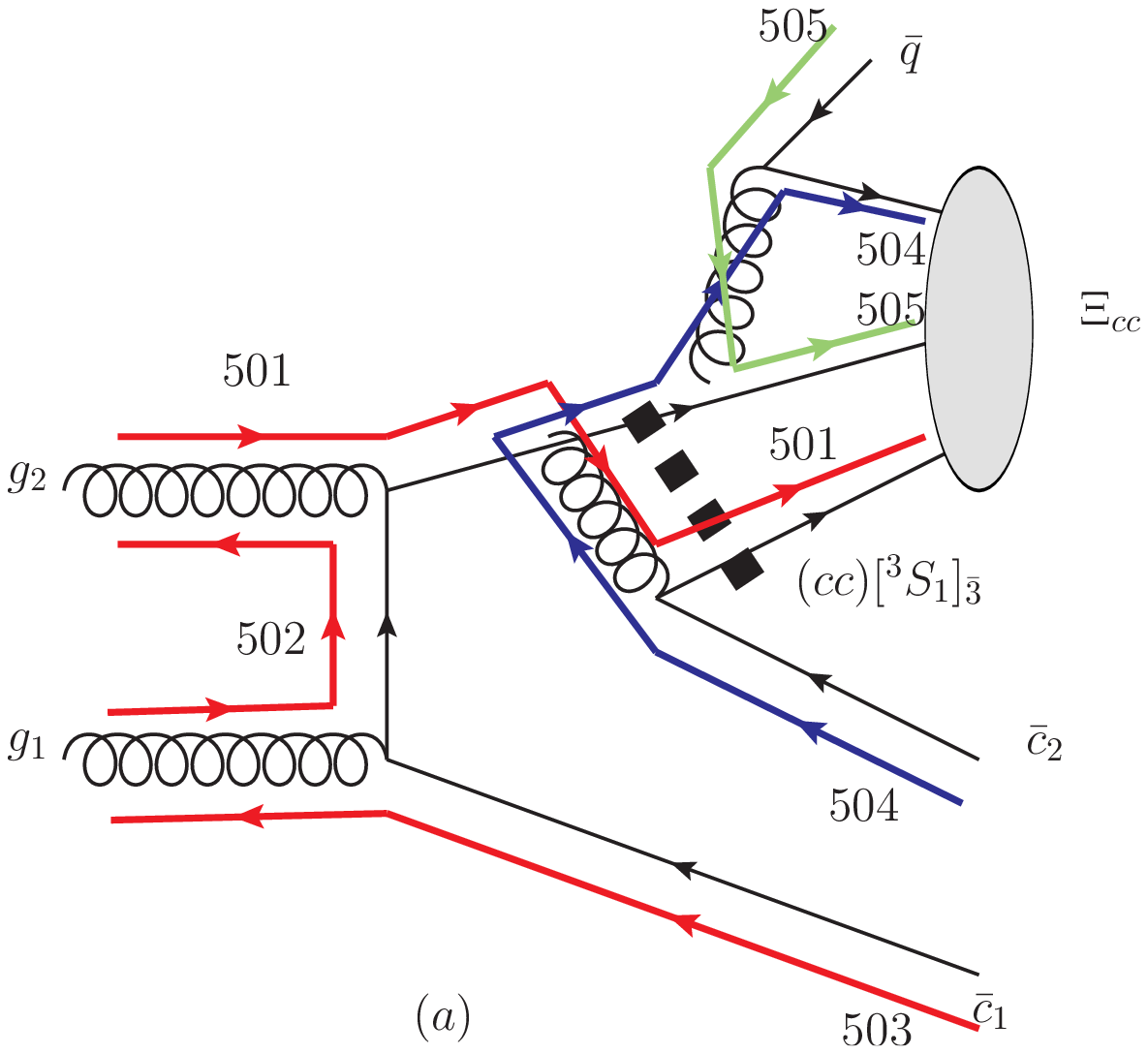}
\includegraphics[width=0.3\textwidth]{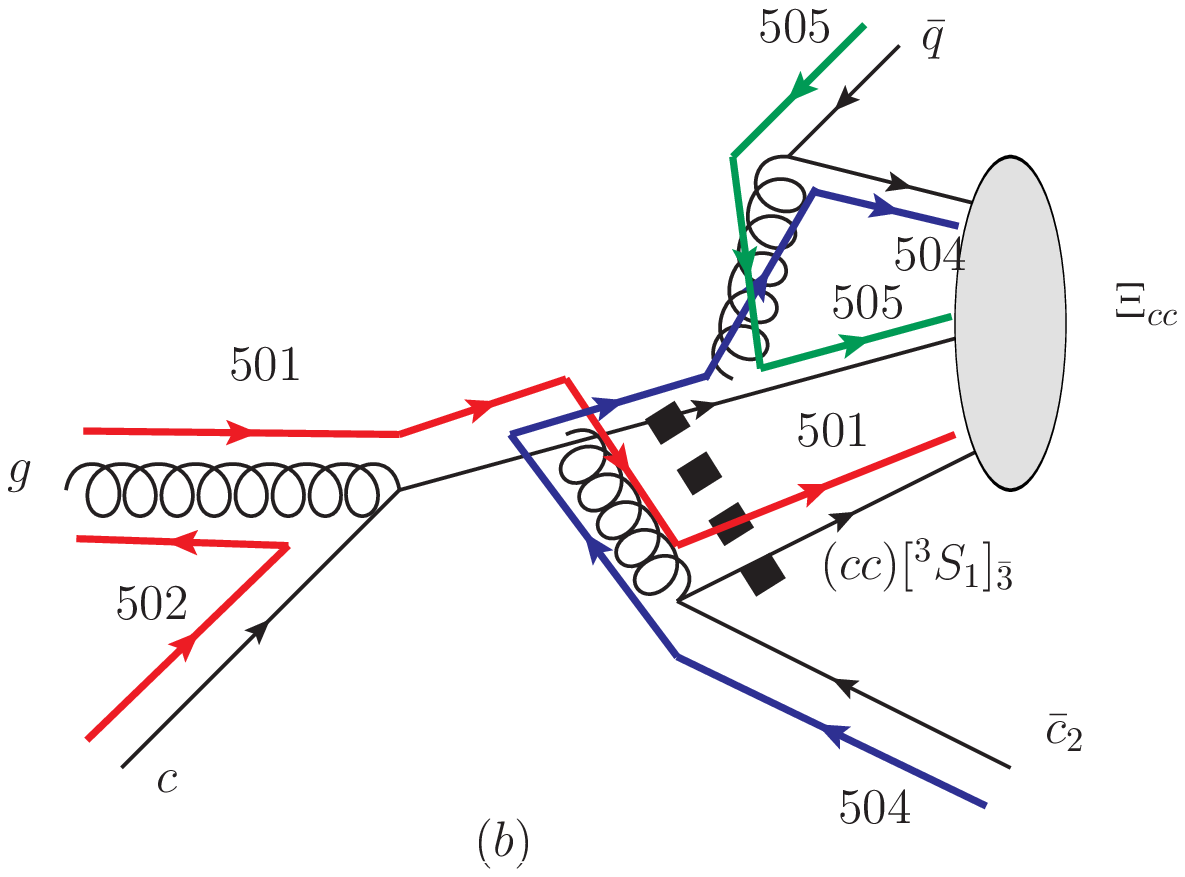}
\includegraphics[width=0.3\textwidth]{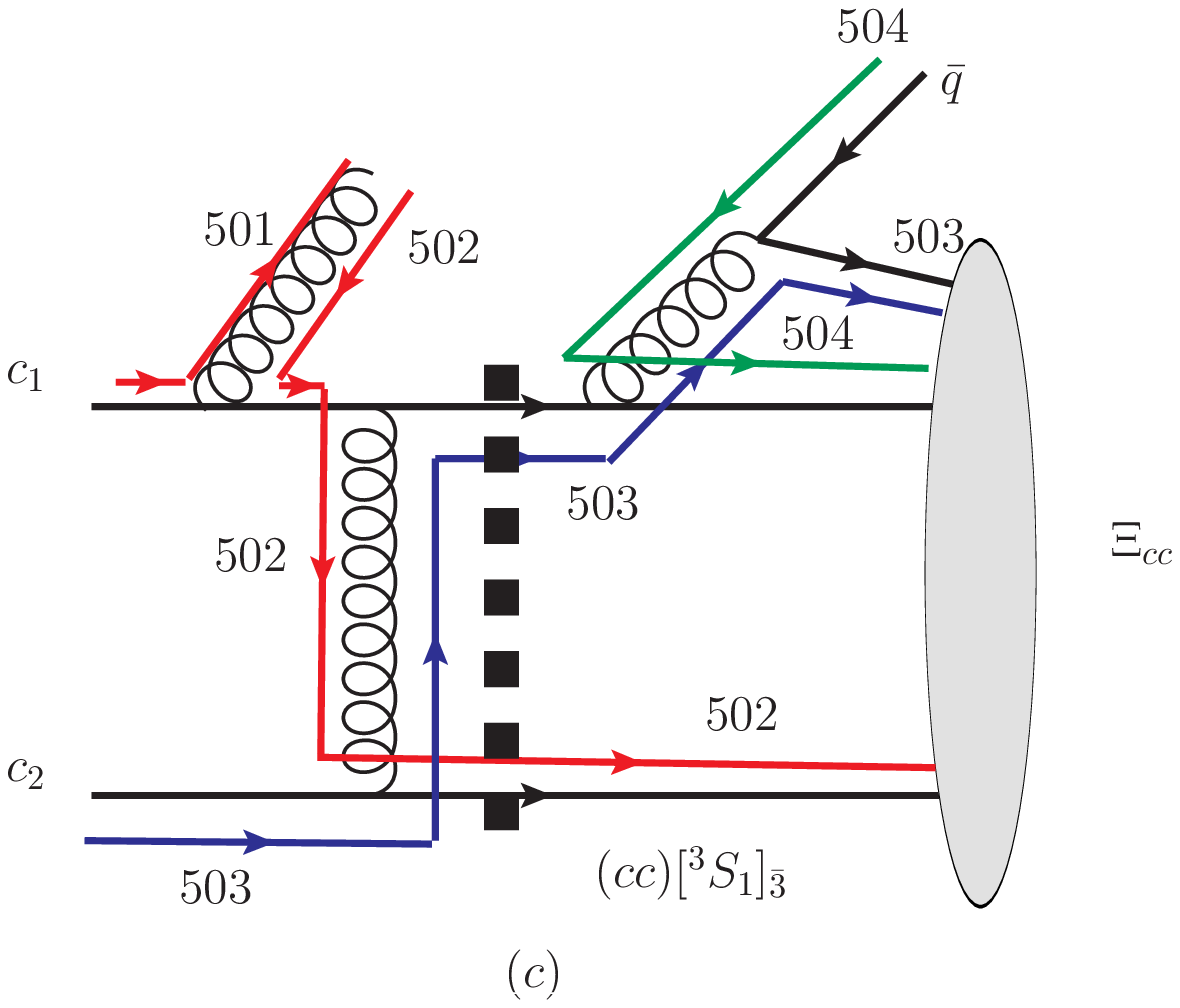}
\caption{(Color on line) Typical color flow for gluon-gluon fusion
mechanism (Left), gluon-charm collision mechanism (Middle) and
charm-charm collision mechanism (Right), where the thick dashed line
shows the corresponding intermediate diquark state
$(cc)[^3S_1]_{\bar{3}}$. Three colorful lines are for color flow
lines according to PYTHIA naming rules. Three colorful lines are for
color flow lines according to PYTHIA naming rules. The black lines
are for Feynman diagrams.} \label{case3s1}
\end{figure}

As for the gluon-gluon fusion mechanism, there are six independent
color factors for Step-A \cite{cqww}: $
C_{1ij}=\frac{1}{2\sqrt{2}}\left(T^aT^b\right)_{mi}G_{mjk}$,
$C_{2ij}=\frac{1}{2\sqrt{2}}\left(T^bT^a\right)_{mi}G_{mjk}$,
$C_{3ij}=\frac{1}{2\sqrt{2}}(T^a)_{mj}(T^b)_{ni}G_{mnk}$,
$C_{4ij}=\frac{1}{2\sqrt{2}}(T^b)_{mj}(T^a)_{ni}G_{mnk}$, $C_{5ij}=
\frac{1}{2\sqrt{2}}\left(T^aT^b\right)_{mj} G_{mik}$ and
$C_{6ij}=\frac{1}{2\sqrt{2}}\left(T^bT^a\right)_{mj} G_{mik}$, where
$i,j=1,2,3$ are color indices of the two outgoing anti-quarks
$\bar{c}$ and $\bar{c}$ respectively, and the indices $a$ and $b$
are color indices for gluon-1 and gluon-2 respectively. Here, the
function $G_{mjk}$ either equals to the anti-symmetric
$\varepsilon_{mjk}$ when the $(cc)$-diquark is in $\bf\bar{3}$
configuration or equals to the symmetric $f_{mjk}$ when the
$(cc)$-diquark is in $\bf 6$ configuration. While implementing
Step-B, one may find that there are only two independent color flows
for the case of $(cc)[^3S_1]_{\bar{3}}$ with $50\%$ probability
each\footnote{This is slightly different from BCVEGPY, the generator
for $B_c$ meson production, where there are five independent color
factors in Step-A which lead to independent color flows accordingly
\cite{bcvegpy}.}. We draw the typical color flows for the
gluon-gluon fusion mechanism, gluon-charm collision mechanism and
charm-charm collision mechanism in FIG.(\ref{case3s1}) respectively,
the other color flow can be obtained by gluon exchange.

Now let us consider the three mechanisms in turn.

For the gluon-gluon fusion mechanism as shown by
FIG.(\ref{case3s1}a), according to the naming rule of PYTHIA, there
are three color flow lines:
\begin{eqnarray}
&& [0,503]\to[502,503]\to[501,502] \to[501,0]\to {\bf colorless\;
bound\; state\; \Xi^{+,++}_{cc}}/\Omega^+_{cc}, \label{ga}\\
&& [0,504]\to [504,0] \to {\bf colorless\; bound\; state\;
\Xi^{+,++}_{cc}}/\Omega^+_{cc}, \label{gb}\\
&& [0,505]\to [505,0] \to{\bf colorless\; bound\; state\;
\Xi^{+,++}_{cc}}/\Omega^+_{cc} ,\label{gc}
\end{eqnarray}
where the final sub-step in Step-B shows that the three partons
$c[501,0]$, $c[505,0]$ and $q[504,0]$ form the Fock state $|ccq>$.

For the gluon-charm collision mechanism, the typical color flow as
shown in FIG.(\ref{case3s1}b) is
\begin{eqnarray}
&& [502,0]\to[501,502] \to[501,0]\to
{\bf colorless\; bound\; state\; \Xi^{+,++}_{cc}}/\Omega^+_{cc},\\
&& [0,504]\to [504,0]\to {\bf colorless\; bound\; state\;
\Xi^{+,++}_{cc}/\Omega^+_{cc}},\\
&& [0,505]\to [505,0]\to {\bf colorless\; bound\; state\;
\Xi^{+,++}_{cc}/\Omega^+_{cc}}.
\end{eqnarray}

For the charm-charm collision mechanism, the typical color flow as
shown by FIG.(\ref{case3s1}c) is
\begin{eqnarray}
&& [501,0]\to[502,501] \to [502,0] \to
{\bf colorless\; bound\; state\; \Xi^{+,++}_{cc}}/\Omega^+_{cc},\\
&& [503,0]\to [503,0] \to {\bf colorless\; bound\; state\; \Xi^{+,++}_{cc}}/\Omega^+_{cc},\\
&& [0,504]\to [504,0]\to{\bf colorless\; bound\; state\;
\Xi^{+,++}_{cc}}/\Omega^+_{cc} .
\end{eqnarray}

\subsection{The color flow for $(cc)[^1S_0]_{6}$ being produced}

\begin{figure}
\centering
\includegraphics[width=0.3\textwidth]{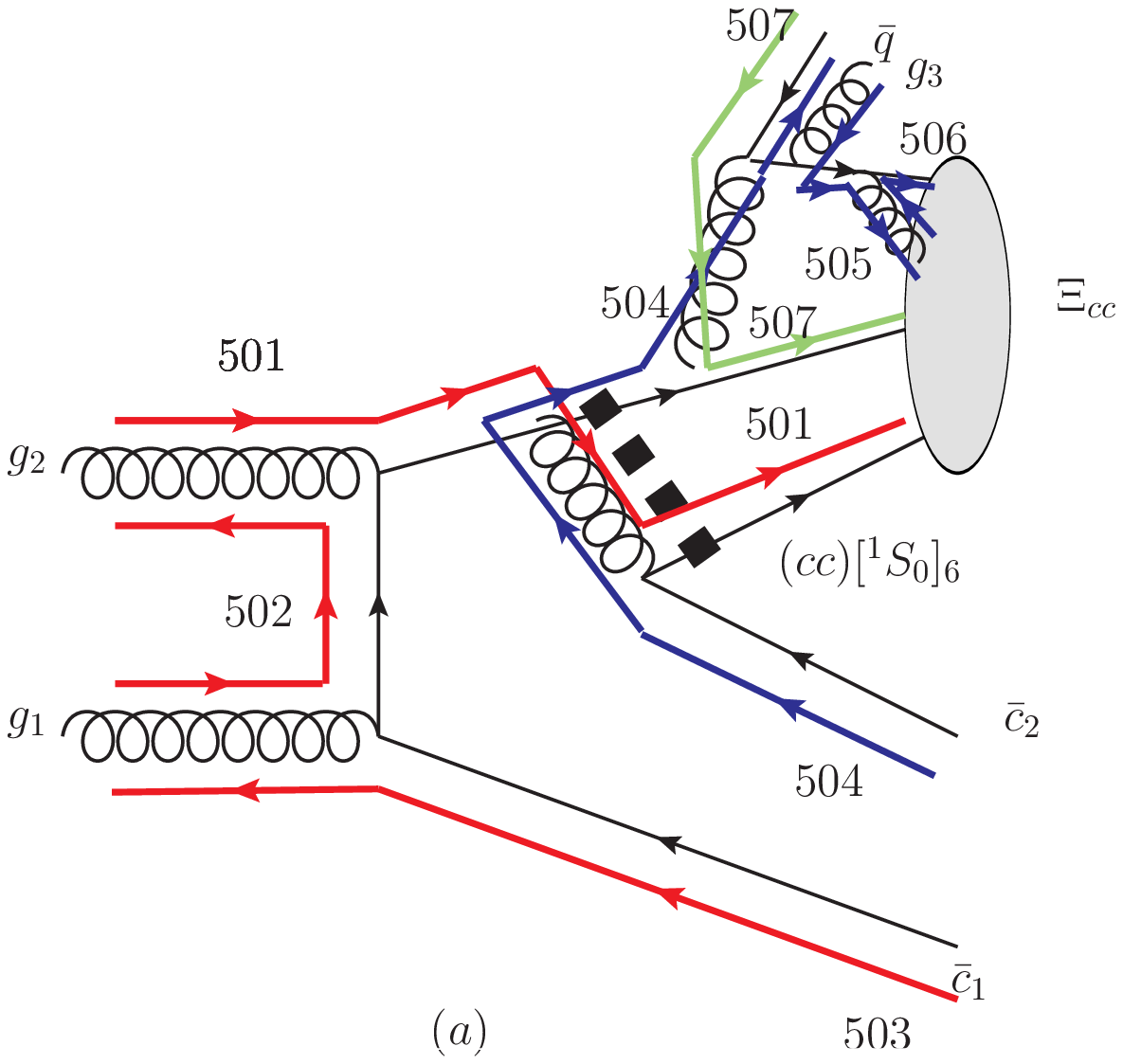}
\includegraphics[width=0.35\textwidth]{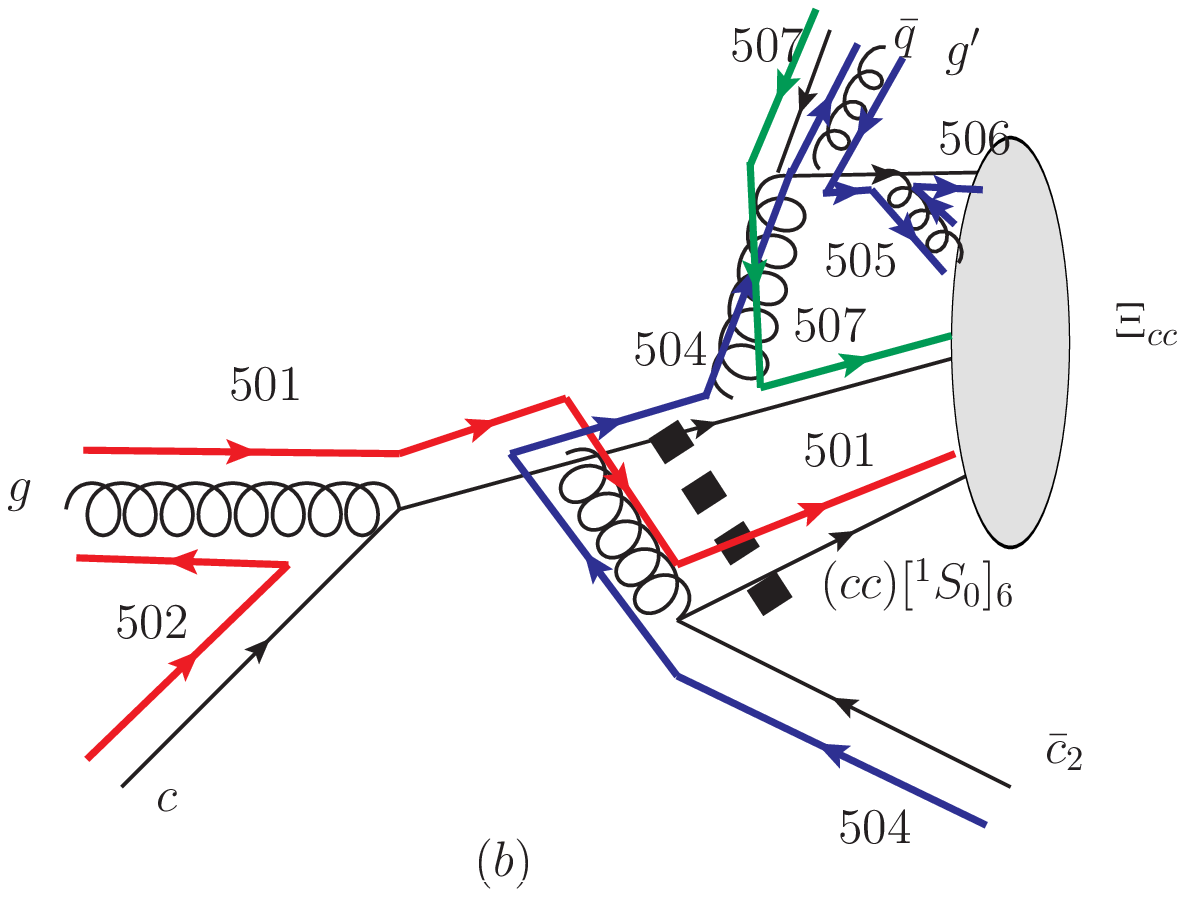}
\includegraphics[width=0.32\textwidth]{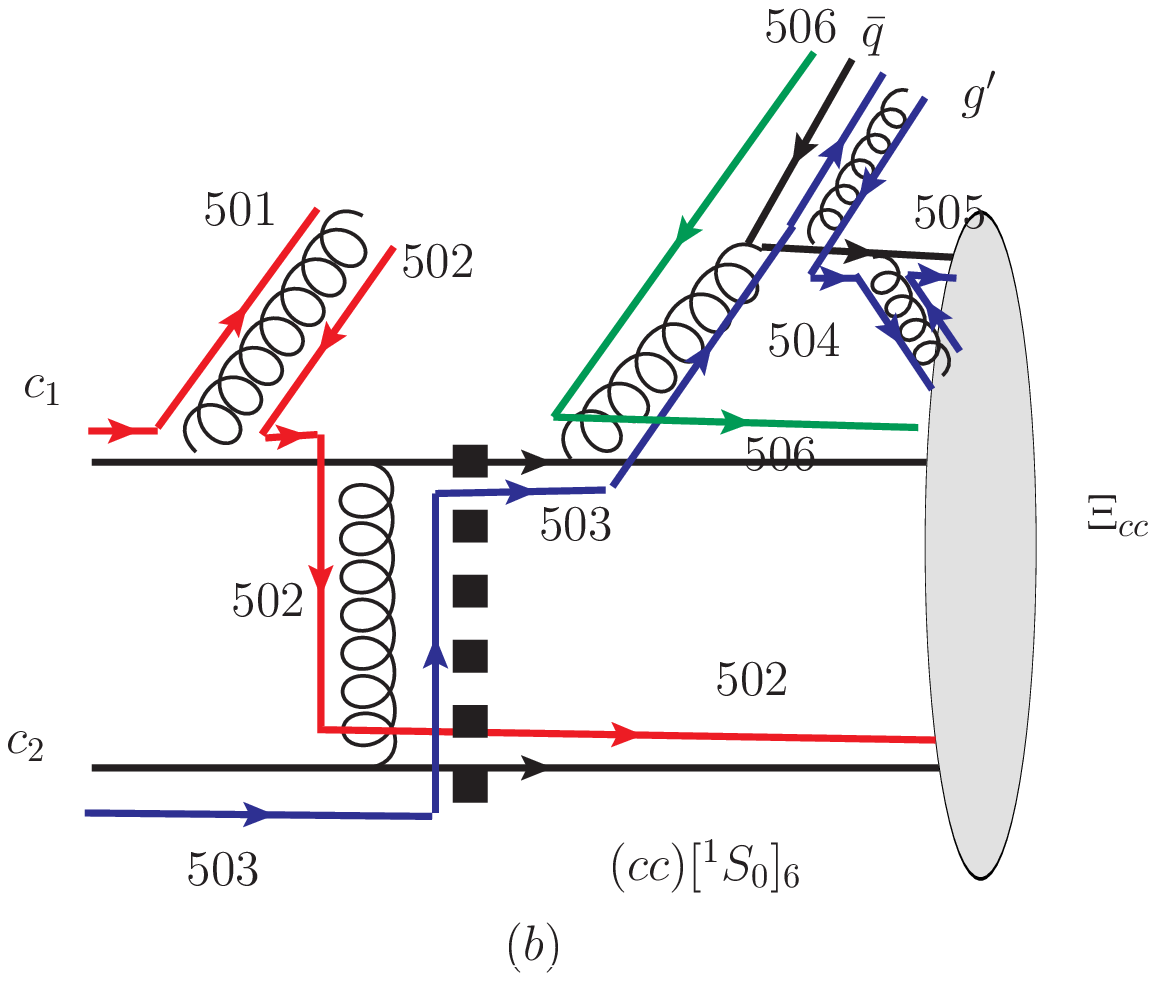}
\caption{(Color on line) Typical color flow for gluon-gluon fusion
mechanism (Left), gluon-charm collision mechanism (Middle) and
charm-charm collision mechanism (Right), where the thick dashed line
shows the corresponding intermediate diquark state
$(cc)[^1S_0]_{6}$. Three colorful lines are for color flow lines
according to PYTHIA naming rules. The black lines are for Feynman
diagrams.} \label{case1s0}
\end{figure}

In the case for $(cc)[^1S_0]_{6}$ diquark being produced, except the
color and the spin of the diquark, programming the Step-A is the
same as that of $(cc)[^3S_1]_{\bar{3}}$, but programming the Step-B
is changed quite a lot:

\begin{itemize}
\item  As pointed out in Ref.\cite{majp}, the fragmentation of
the color-sixtet diquark into the baryon is most likely via the Fock
state $|ccqg\rangle$ instead via the Fock state $|ccq\rangle$ (the
later need to change the heavy diquark spin, that it must be
suppressed some). Note that as pointed out above, for the production
of $\Xi_{bc}$ without `identity particle rule', the role of the two
Fock states $|bcq\rangle$ and $|bcqg\rangle$ played in the
fragmentation is similar. So the difference in color flow for the
$|bcq\rangle$ and $|bcqg\rangle$ does not cause different result in
$\Xi_{bc}$ production.
\item Since the diquark $(cc)[^1S_0]_{6}$ is in a color sixtet, in order
to keep color flow of the production correctly, at least one more
extra soft gluon, besides a light anti-quark, must be remained in
the final state when the fragmentation is completed. It is because
that, as the most simple case, a gluon and an anti-quark can
construct a color-sixtet:
\begin{equation}
8 \bigotimes \bar{3}=15 \bigoplus {\bf 6} \bigoplus \bar{3}\,,
\end{equation}
and the sixtet may fully `absorb' the color of the diquark
$(cc)[^1S_0]_{6}$. Namely it is only due to the request to make
color flow correctly that here an extra gluon is needed.
\item The most simple `mechanism', which may be the biggest one, to
generate the extra soft gluon is by emitting from another soft light
quark or a soft gluon, which is necessary in the fragmentation
sub-process (one typical example can be found in
FIG.(\ref{case1s0})), so the extra so soft gluon causes negligible
consequence in momentum distribution, even though all the subprocess
are of non-perturbative nature. Hence in dealing with the momentum
distribution among the baryon and jets etc, we can treat the soft
anti-quark and soft gluon in the final state as a whole object
\footnote{It is reasonable, since numerically, one find that the
soft quark itself gives negligible contributions to the momentum
distributions for the rest particles in final state.}. Once more one
may see that to introduce the soft anti-quarks and so soft gluons is
only for keeping the color conserves (flows correctly) in the
production.
\end{itemize}

Some of typical color flows in the production via $(cc)[^1S_0]_{6}$
are shown in FIG.(\ref{case1s0}), where FIG.(\ref{case1s0}a),
FIG.(\ref{case1s0}b) and FIG.(\ref{case1s0}c) are for gluon-gluon
fusion, gluon-charm collision and charm-charm collision mechanisms
respectively. All possible colors for the production may be obtained
by interchanging the connect points among the soft partons properly.

For the gluon-gluon fusion mechanism, i.e. the case as shown by
FIG.(\ref{case1s0}a), the three color flow lines are:
\begin{eqnarray}
&a).\;\;\; &[0,503]\to[502,503]\to[501,502] \to[501,0] \nonumber \\
&& \to
{\bf colorless\; bound\; state\; \Xi^{+,++}_{cc}}/\Omega^+_{cc}, \\
&b).\;\;\;&[0,504]\to [504,505]\to [505,506]
\overrightarrow{\to [506,0]\to}  \nonumber \\
&&{\bf colorless\;
bound\; state\; \Xi^{+,++}_{cc}}/\Omega^+_{cc}, \\
&c).\;\;\;& [0,507]\to [507,0] \to{\bf colorless\; bound\; state\;
\Xi^{+,++}_{cc}}/\Omega^+_{cc},
\end{eqnarray}
where an extra soft gluon with color $[505,506]$ is introduced, in
comparison with the case of $(cc)[^3S_1]_{\bar{3}}$, and connected
to the second color flow line respectively. For the second
color-flow line Eq.(15), both the gluon color $g[505,506]$ and the
quark color $q[506,0]$ flow into the baryon. Thus here the equations
Eqs.(14-16) correspond to that the four partons $c[501,0]$,
$c[507,0]$, $q[506,0]$ and $g[505,506]$ form the Fock state $|ccqg>$
finally. It can be found that the color octet gluon $g_3[504,505]$
and the color anti-triplet anti-quark $\bar{q}[0,507]$ can form a
color ${\bf 6}$ state so as to keep the color conservation of the
process.

For the gluon-charm collision mechanism, as shown by
FIG.(\ref{case1s0}b), we have
\begin{eqnarray}
&a).\;\;\; & [502,0]\to[501,502] \to[501,0]\to
{\bf colorless\; bound\; state\; \Xi^{+,++}_{cc}}/\Omega^+_{cc},\\
&b).\;\;\; & [0,504]\to [504,505]\to [505,506] \overrightarrow{\to
[506,0]\to}\nonumber\\
&&{\bf colorless\; bound\;state\; \Xi^{+,++}_{cc}/\Omega^+_{cc}},\\
&c).\;\;\; & [0,507]\to [507,0]\to {\bf colorless\; bound\; state\;
\Xi^{+,++}_{cc}/\Omega^+_{cc}} .
\end{eqnarray}
The equations Eqs.(17-19) correspond to the four partons $c[501,0]$,
$c[507,0]$, $q[506,0]$ and $g[505,506]$ form the Fock state $|ccqg>$
finally.

For the charm-charm collision mechanism, as shown by
FIG.(\ref{case1s0}c), we have
\begin{eqnarray}
&a).\;\;\; & [501,0]\to[502,501] \to [502,0]\to
{\bf colorless\; bound\; state\; \Xi^{+,++}_{cc}}/\Omega^+_{cc},\\
&b).\;\;\; & [503,0]\to [503,504]\to [504,505] \overrightarrow{\to
[505,0] \to} \nonumber\\
&&{\bf colorless\;
bound\; state\; \Xi^{+,++}_{cc}}/\Omega^+_{cc},\\
&c).\;\;\; & [0,506]\to [506,0]\to{\bf colorless\; bound\; state\;
\Xi^{+,++}_{cc}}/\Omega^+_{cc} .
\end{eqnarray}
The equations Eqs.(20-22) correspond to the four partons $c[502,0]$,
$c[506,0]$, $q[505,0]$ and $g[504,505]$ form the Fock state $|ccqg>$
finally. In Eqs.(15,18,21) the symbol `` $\overrightarrow{\to
[\cdots]\to}$ '' means a soft gluon is accompanied.

\subsection{The color flow lines ended at a baryon and
simplification with `cheat'}

In the above two subsection, based on the color-flow decomposition
method of Ref.\cite{color} we have designed a strict way to deal
with the color flows for the baryon production.

In fact since a baryon contains three valance quarks in different
colors so as to form a colorless object, thus under the color-flow
decomposition method, there must be three color flow lines being
ended at a baryon. Namely it is different from the case of meson,
where the color flow lines of the quark and anti-quark inside a
meson are continued. To simplify the present situation, reminding
the fact that $3 \bigotimes 3=6 \bigoplus {\bf \bar{3}}$ in general
QCD SU(3) color space: 1) Firstly, we can combine any two of the
flow lines ended with two quarks into one anti-flow line ended with
one anti-quark with a color $\bar{3}$ (the direction of the line is
turned back completely) that is different from the two quarks (the
third kind of color in respect those kinds of the two quarks); 2)
Secondly, such anti-flow line obtained by the combination may be
continued (connected) to the third color-flow line in the baryon; 3)
Finally, as a consequence, the color-flow lines ended at a baryon
become `joined without ends' at all, which is the requirement of the
color-singlet bound state.

With the method for the color-flow lines ended at baryon,
practically, the present version of PYTHIA (version 6.4) can
generate full events with two or less independent color-flow lines
in most cases well even a baryon being involved.

When there are three or more independent color-flow lines in
process, PYTHIA will stop at the step of running the `final parton
showers' or parton cascade radiations etc, and will present an error
message to show that the color flow rearrangement is wrong during
the parton's evolution process. For the present problem to generate
the full events for the baryon production, it is found that in
GENXICC all the mentioned production mechanisms: gluon-gluon fusion,
gluon-charm collision and charm-charm collision, have exactly three
independent color flows as shown by the last subsections when the
color flow lines ended at the produced baryon are not treated as the
above (to combine two of the three lines into one anti-color flow
line). So with GENXICC1.0 being directly complemented, PYTHIA can
not work well with these mechanisms at the step of `final parton
showers' or parton cascade radiations etc. To overcome the
difficulty and to achieve full useful information for each event of
the baryon production, the best approach perhaps is that PYTHIA
itself is improved to deal with three or more color-flow lines'
processes etc, e.g. to add the useful intermediate diquark states
with color ${\bf\bar{3}}$ or color ${\bf 6}$, such as
$(cc)[^3S_1]_{\bar{3}}$, $(cc)[^1S_0]_{6}$ etc, into the particle
tabular of PYTHIA, and properly to treat the color-flow lines ended
at the baryon in fragmenting them into baryons accordingly. We
suspect that it can be realized as done in a similar way as that in
PYTHIA \cite{pythia} to treat color-octet mechanisms for the
$J/\Psi$ production without difficulty.

Since GENXICC1.0, the previous version of the generator, without
precisely strict color-flow lines being defined, cannot run well for
the available versions of PYTHIA, and furthermore, now there is no
the improved PYTHIA as mentioned above, thus in GENXICC2.0 the
upgraded version of the generator, we have to suggest another way so
as to complete the full-event generation under a controlled
approximate level (not to deviate from true too much) with the color
flow lines ended at a baryon. Namely, we properly vary the color
flows of the process as depicted in the last subsections, so as to
make PYTHIA running well for final parton shower process. Aiming at
the purpose, we precisely restrict the state constructed by either
$(ccq)$ or $(ccqg)$ is equal to the desired colorless baryon, then
I) to keep color conservation, all the outgoing particles' colors
are arranged properly such that they form a color state as that of
the incoming particles; II) The details in color of the constitute
partons are ignored. PYTHIA is cheated to believe that the state
constructed by either $(ccq)$ or $(ccqg)$ is colorless baryon, so we
force it `steady' without any evolution, thus any programs, which
are called from PYTHIA, may run smoothly. In this way and only to
introduce necessary soft enough light quark(s) anti-quark(s) and
gluon(s), the momentum distribution and the other factors for the
production will not be affected much. Moreover, we further simplify
the program for the production in color as follows: 1) Ignoring the
details about the color flow when the diquark fragments into a
colorless baryon, the initial color-flow lines are reduced to be
less than or equal to two. More explicitly, for instance the three
color-flow lines in Eqs.(\ref{ga},\ref{gb},\ref{gc}) for the
gluon-gluon fusion mechanism ($(cc)[^3S_1]_{\bar{3}}$) lead to two
independent color-flow lines: $[0,503]\to[502,503]\to[501,502]
\to[501,0]\to {\bf colorless\; bound\; state}$ and $[0,504]\to
[504,0]\to {\bf colorless\; bound\; state}\to [0,504]$; 2) A working
assumption in program is made by setting the outgoing light
anti-quark's color to be the same as that of the outgoing
$\bar{c}_2$ or $(\bar{c})$ (see FIG.\ref{case3s1}) in the case of
gluon-gluon fusion or gluon-charm collision mechanism respectively;
3) In the production, to `introduce' a soft anti-quarks and a soft
gluon for the diquark $(cc)[^1S_0]_{6}$ is only to keep the color
conservation, so we may formally combine these two soft objects as
an effective soft `parton'. Therefore, as for the color
freedom-degree only, for the $(cc)[^1S_0]_{6}$ we force the outgoing
soft gluon and soft anti-quark to be combined into an effective one
color-sixtet object (parton) formally, so the color flow for both
$(cc)[^3S_1]_{\bar{3}}$ and $(cc)[^1S_0]_{6}$ is treated in the same
manner in the program. Namely in color, the effective soft `parton'
is $\bar{q}$ for $(cc)[^3S_1]_{\bar{3}}$ and is $(\bar{q}g)$ for
$(cc)[^1S_0]_{6}$ accordingly. Moreover, in GENXICC2.0, we provide
an interface of these soft objects to PYTHIA and users, who are
interested in this part, can conveniently use these code to do the
event evolution. We should note here that this simplification in
treating the color flows cannot be applied to the charm-charm
collision mechanism, i.e., we can not simultaneously simplify the
three independent color-flow lines into two and also keep the
process in color conservation. However, the contribution from this
mechanism is small in comparison with the other two mechanisms at
higher energy colliders as TEVATRON and LHC\cite{cqww} fortunately,
so in the generator we safely ignore the charm-charm collision
mechanism at all.

\section{Discussion}
To generate full events for the production of the double heavy
baryons with GENXICC complemented into PYTHIA, as pointed out above,
obviously the best way is to improve PITHIA with proper treatment
the color-flow lines ended at the baryon. Whereas due to the fact
that the improved version of PYTHIA is not available now, thus we
`have to suggest' another way with certain approximation, i.e. when
writing down the program, the upgraded version has additionally made
certain simplification with `cheat' so as to generate the full
events by applying GENXICC2.0, the upgraded generator, being
complemented in PYTHIA, quite efficiently. However we should point
out here that due to approximation and simplification with `cheat'
explained in the above subsection, the obtained information about
the 'tiny jets', corresponding to the soft anti-quark and soft
gluon(s) produced in fragmentation of double heavy diquark, may not
be very reliable, although the information about hard jets and the
baryon as well in the full events is reliable.

\vspace{1cm}

{\bf Acknowledgments}: The authors would like to thank Braden Keim
Abbott and Richard J. Van Kooten for drawing their attention to the
program defaulting the precise fragmentation from the heavy diquarks
into the baryons in GENXICC1.0. This work was supported in part by
Natural Science Foundation of China (NSFC) under Grant No.10805082
No.10875155, No. 10847001 and No. 10875155, and by Natural Science
Foundation Project of CQ CSTC under Grant No.2008BB0298. This
research was also supported in part by the Project of Knowledge
Innovation Program (PKIP) of Chinese Academy of Sciences, Grant No.
KJCX2.YW.W10.


\begin{thebibliography}{s2}

\bibitem{D0} Private communications with Braden Keim Abbott and Richard J. Van
Kooten (D0 Collaboration).

\bibitem{genxicc} Chao-Hsi Chang, Jian-Xiong Wang and Xing-Gang Wu,
Comput.Phys. Commun.{\bf 177}, 467(2007).

\bibitem{cqww} Chao-Hsi Chang, Cong-Feng Qiao, Jian-Xiong Wang and Xing-Gang
Wu, Phys. Rev. D{\bf 73}, 094022(2006).

\bibitem{cmqw} Chao-Hsi Chang, Jian-Ping Ma, Cong-Feng Qiao and Xing-Gang
Wu, J.Phys. G{\bf 34}, 845(2007).

\bibitem{pythia} T. Sjostrand, S. Mrenna and P. Skands, JHEP {\bf 0605},
026(2006).

\bibitem{rambo} R. Kleiss and W.J. Stirling, Comput. Phys. Commun. {\bf
40}, 359(1986).

\bibitem{vegas} G.P. Lepage, J. Comp. Phys {\bf 27}, 192 (1978).

\bibitem{majp} J.P. Ma and Z.G. Si, Phys. Lett. B{\bf 568},
135(2003).

\bibitem{bcvegpy} Chao-Hsi Chang, Chafik Driouich, Paula Eerola and Xing-Gang Wu,
Comput.Phys. Commun. {\bf 159}, 192 (2004); Chao-Hsi Chang,
Jian-Xiong Wang and Xing-Gang Wu, Comput.Phys. Commun. {\bf 174},
241 (2006); Chao-Hsi Chang, Jian-Xiong Wang and Xing-Gang Wu,
Comput.Phys. Commun. {\bf 175}, 624 (2006).

\bibitem{color} F. Maltoni, K. Paul, T. Stelzer, S. Willenbrock,
Phys.Rev. D{\bf 67}, 014026(2003).

\end{thebibliography}
\end{document}